\documentstyle[12pt]{article}
\begin{document}
\begin{center}
{\large On the Monge-Amp\`ere equivalent of the sine-Gordon equation}\\[15mm]
{\large E. V. Ferapontov$^*$  } and {\large Y. Nutku}\\[10mm]
 T\"{U}B\.{I}TAK - Marmara Research Center \\
 Research Institute for Basic Sciences \\
 Department of Physics \\
 41470 Gebze, Turkey \\[50mm]
\end{center}
\noindent

   Surfaces of constant negative curvature in Euclidean space
can be described by either the sine-Gordon equation
for the angle between asymptotic directions, or a Monge-Amp\`ere
equation for the graph of the surface.
We present the explicit form of the correspondence between these two
integrable non-linear partial differential equations using their
well-known properties in differential geometry.
We find that the cotangent of the angle between asymptotic directions
is directly related to the mean curvature of the surface.
This is a B\"{a}cklund-type transformation between the sine-Gordon
and Monge-Amp\`ere  equations.

\vspace{1 cm}

\noindent
----------------------------------------------------------------------------
\noindent

$^*$ permanent address: Institute for Mathematical Modelling,
Academy of Science of Russia, Miusskaya 4, Moscow 125047, Russia

\pagebreak

\section{Introduction}

    Monge-Amp\`ere equations have so far been excluded from extensive
searches for integrable non-linear evolution equations \cite{z}. However,
their origin lies in geometry and quite frequently
we find them in the same category of problems where other well-known
integrable systems also find their natural setting.
Some of these Monge-Amp\`ere equations were listed in \cite{ns}
but the identification of the Monge-Amp\`ere equivalent of completely
integrable non-linear evolution equations is still an outstanding problem
in many interesting cases. Perhaps the most remarkable of all such
correspondences was established by J\"{o}rgens \cite{j} who showed that
the elliptic Monge-Amp\`ere equation with one on right hand side
is equivalent to the equation governing minimal surfaces.
For the hyperbolic Monge-Amp\`ere equation the corresponding equation is
the Born-Infeld equation, or the Euler equations for Chaplygin gas \cite{mn}.

   We shall now present the Monge-Amp\`ere equivalent of the sine-Gordon
equation. It is well-known that
\begin{equation}   \begin{array}{rcl}
u_{xx} u_{yy} - u_{xy}^{\;\;\;2} & = & - \, K^2 \, , \\[2mm]
K &  \equiv  & 1 + u_{x}^2 + u_{y}^2
\end{array}
\label{ma}
\end{equation}
where $z = u(x,y)$ is the graph of the surface, and the sine-Gordon equation
\begin{equation}
\phi_{\xi \eta} = sin \, \phi
\label{sg}
\end{equation}
where $\phi(\xi, \eta)$ is the angle between asymptotic directions
must be related as they both govern surfaces of constant negative curvature.
This observation occupies a central role in Anderson and Ibragimov's
\cite{ai} exposition of the Bianchi-Lie transformation itself, as in their
discussion they revert to either one of eqs.(\ref{ma}), or (\ref{sg}) freely
even without presenting an explicit transformation between them.
We shall now show that in the geometrical context
the relation between $\phi$ and $u$ can be obtained directly.
The result, given by eq.(\ref{ctanphi}) below, is an expression for
the cotangent of the angle between asymptotic directions
in terms of the mean curvature of the surface.

\section{Geometrical preliminaries}

    We recall the first and second fundamental forms of a surface in $E^3$
with curvature $-1$ for which the Gauss-Codazzi equations give rise to the
Monge-Amp\`ere (\ref{ma}) and sine-Gordon (\ref{sg}) equations respectively.
For the Monge-Amp\`ere equation they are given by
\begin{equation}                  \begin{array}{lll}
 d s_{1}^{\;2} & =  &
 ( 1 + u_{x}^2 ) \, d x^2 + 2 u_{x} u_{y}  \, d x  d y
  + ( 1 + u_{y}^2 )  \, d y^2 \, , \\[2mm]
 d s_{2}^{\;2} & =  &
 \frac{\textstyle{1}}{\textstyle{ \sqrt K }}
\left(    u_{xx} \, d x^2 + 2 u_{xy} \, d x d y
                 + u_{yy} \, d y^2 \right) \, ,
\end{array}
\label{12ma}
\end{equation}
while for the sine-Gordon case we have
\begin{equation}                  \begin{array}{lll}
 d s_{1}^{\;2} & =  & d \xi^2 + 2 cos \phi \, d \xi  d \eta + d \eta^2 \, ,
 \\[2mm]
 d s_{2}^{\;2} & =  &  2 \phi_{\xi \eta} \, d \xi d \eta \, .
\end{array}
\label{12sg}
\end{equation}
In order to establish the equivalence between eqs.(\ref{ma}) and (\ref{sg})
we shall require that their first and second fundamental forms must agree.

   To this end we note that $\xi, \eta$ entering into the sine-Gordon
equation are the asymptotic coordinates on the surface and we must first
find their counterparts in the case of the Monge-Amp\`ere equation.
The characteristics of eq.(\ref{ma}) satisfy
\begin{equation}
u_{xx} {x'}^2 + 2 u_{xy} x' y' + u_{yy} {y'}^2 = 0
\label{charac}
\end{equation}
which can be written in the form
\begin{equation}
[ u_{xx} x' + ( u_{xy} + K ) y' ] [  ( u_{xy} + K ) x' + u_{yy} y' ] = 0
\label{characp}
\end{equation}
and therefore we can introduce $\xi, \eta$ such that
\begin{equation}                  \begin{array}{lll}
 d \xi & =  &   A  \left[ u_{xx} d x +   ( u_{xy} + K )   d y \right]
\\[1mm]
 d \eta & = &   B  \left[ ( u_{xy} + K )   d x +  u_{yy} d y  \right]
\end{array}
\label{xieta}
\end{equation}
where $A, B$ may depend on $u$ and its derivatives.
The requirement that the first fundamental forms in eqs.(\ref{12ma})
and (\ref{12sg}) must agree yields
\begin{equation}                  \begin{array}{rll}
A^2 & =  &
 \frac{ \textstyle{  ( 1 + u_{x}^2 ) u_{yy}^{\;\;2} - 2 u_{x} u_{y} u_{yy}
 ( u_{xy} + K )   + ( 1 + u_{y}^2 ) ( u_{xy} + K )^2  } }
 { \textstyle{ 4 K^2 ( u_{xy} + K )^2 }}         \\[4mm]
B^2 & = &
 \frac{ \textstyle{  ( 1 + u_{y}^2 ) u_{xx}^{\;\;2} - 2 u_{x} u_{y} u_{xx}
 ( u_{xy} + K )   + ( 1 + u_{x}^2 ) ( u_{xy} + K )^2 } }
 { \textstyle{ 4 K^2 ( u_{xy} + K )^2 }}         \\[4mm]
A B \, cos \phi & = &
-  \frac{ \textstyle{  ( 1 + u_{y}^2 ) u_{xx} - 2 u_{x} u_{y} u_{xy}
  + ( 1 + u_{x}^2 ) u_{yy} } }
 { \textstyle{ 4 K^2 ( u_{xy} + K ) }}
\end{array}
\label{abc}
\end{equation}
and these relations are sufficient to establish the identity
of the second fundamental forms as well, {\it c.f.} also
eq.(\ref{ctanphi}) below.

   The exterior derivative of eqs.(\ref{xieta}) must vanish. Hence it follows
that the coefficients of the 1-forms on the right hand side of these equations
are conserved quantities for the Monge-Amp\`ere equation (\ref{ma}).
This can be verified by a straight forward but lengthy calculation
which also serves to establish the local existence of $\xi$ and $\eta$.

\section{The correspondence}

    Eqs.(\ref{abc}) which insure the agreement of the first and second
fundamental forms for the Monge-Amp\`ere and the sine-Gordon equations
result in a B\"{a}cklund-type transformation between $ \phi$ and $u$.
This is most conveniently expressed in the form
\begin{equation}
ctn \, \phi = -  \frac{1}{ 2 \, K^{3/2} } \, \left[
 ( 1 + u_{y}^2 ) \, u_{xx} - 2 u_{x} u_{y} \, u_{xy}
  + ( 1 + u_{x}^2 ) \, u_{yy}  \right]
\label{ctanphi}
\end{equation}
which is simply the mean curvature of the surface obtained by the
contraction of the first and second fundamental forms in eqs.(\ref{12ma}).
Remembering that $\sqrt{K}$ is the coefficient of the volume element
which also serves as the Lagrangian for minimal surfaces we can also write
\begin{equation}
ctn \, \phi =  \frac{1}{2}  \, \frac{\delta }{\delta u} \; \sqrt{K}
\label{log}
\end{equation}
where $\delta$ denotes the variational derivative.
But the easiest way of remembering the result (\ref{ctanphi}) is through
its geometrical meaning, namely,
the cotangent of the angle between asymptotic directions
is minus one half the mean curvature of the surface.

\section{Remarks}

   In the discussion of the existence of $\xi, \eta$ through
eqs.(\ref{xieta}) we found integrals of the Monge-Amp\`ere
equation (\ref{ma}) that vanish along the characteristics.
In general these are called generalized flow functions,
or characteristic integrals. Vessiot \cite{vessiot}
has presented a complete description of Monge-Amp\`ere equations
which admit infinitely many generalized flow functions.
This list coincides with Zvyagin's list \cite{zvyagin} of second order
equations reducible to the wave equation by a B\"{a}cklund transformation
of $B_3$-type in the terminology of Goursat.
In our case there are only two generalized flow functions, one
along each characteristic. We also note that Mukminov \cite{muk}
considered transformations of Monge-Amp\`ere equations to
characteristic form using generalized flow functions which depend
only on first derivatives. The generalized flow functions in
eqs.(\ref{xieta}) differ from Mukminov's through their
dependence on second derivatives.

\section{Conclusion}

   We have shown that there is a B\"{a}cklund-type transformation
between the Monge-Amp\`ere equation and the sine-Gordon equation which
relates the cotangent of the angle between asymptotic directions to the
mean curvature of the surface. This result is sufficient to
translate the infinite hierarchy of conserved quantities, auto-B\"{a}cklund
transformation and other well-known properties of the sine-Gor\-don
equation (\ref{sg}) to its Monge-Amp\`ere equivalent (\ref{ma}).

\vspace{10mm}

\noindent
{\it Acknowledgements}
\vspace{4mm}

   We thank T\"{U}B\.{I}TAK for instituting a program of visits by scientists
from the former Soviet Union which made this collaboration possible.

\end{document}